\documentclass[superscriptaddress,showpacs,preprintnumbers,amsmath,amssymb,nofootinbib]{revtex4}

\usepackage{graphicx}
\usepackage{dcolumn}
\usepackage{bm}

\begin{document}

\preprint{}

\title{Curvature effect on nuclear ``pasta'': Is it helpful for gyroid appearance?}

\author{Ken'ichiro Nakazato}
 \email{nakazato@rs.tus.ac.jp}
 \affiliation{Department of Physics, Faculty of Science \& Technology, Tokyo University of Science, Yamazaki 2641, Noda, Chiba 278-8510, Japan
}%

\author{Kei Iida}
 \affiliation{Department of Natural Science, Kochi University, 2-5-1 Akebono-cho, Kochi 780-8520, Japan
}%

\author{Kazuhiro Oyamatsu}
 \affiliation{Department of Human Informatics, Aichi Shukutoku University, Nagakute-Katahira 9, Nagakute, Aichi 480-1197, Japan
}%

\date{\today}

\begin{abstract}
In supernova cores and neutron star crusts, nuclei are thought to deform to rodlike and slablike shapes, which are often called nuclear pasta. We study the equilibrium properties of the nuclear pasta by using a liquid drop model with curvature corrections. It is confirmed that the curvature effect acts to lower the transition densities between different shapes. We also examine the gyroid structure, which was recently suggested as a different type of nuclear pasta by analogy with the polymer systems. The gyroid structure investigated in this paper is approximately formulated as an extension of the periodic minimal surface whose mean curvature vanishes. In contrast to our expectations, we find from the present approximate formulation that the curvature corrections act to slightly disfavor the appearance of the gyroid structure. By comparing the energy corrections in the gyroid phase and the hypothetical phases composed of $d$-dimensional spheres, where $d$ is a general dimensionality, we show that the gyroid is unlikely to belong to a family of the generalized dimensional spheres.
\end{abstract}



\maketitle

\section{Introduction} \label{intro}
\begin{figure}
\begin{center}
\includegraphics[height=76mm]{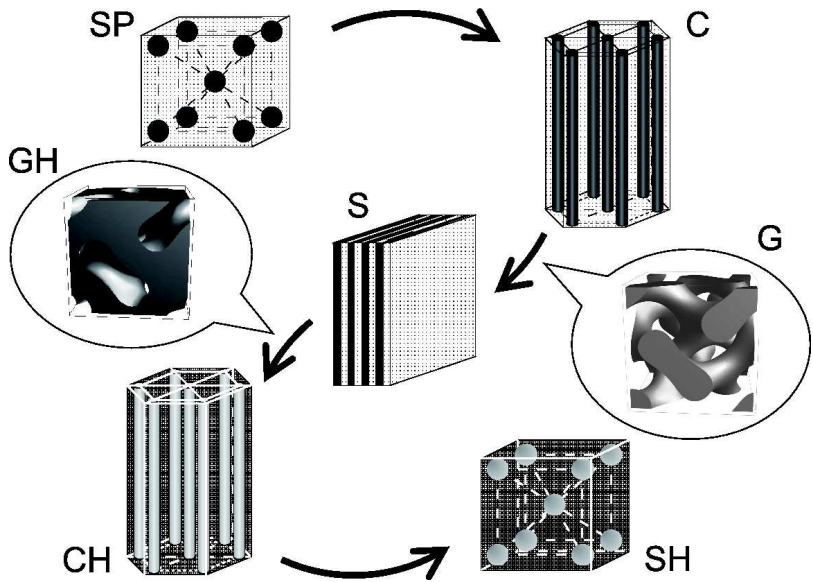}
\caption{Candidates for equilibrium nuclear shape. In this figure, the following notation is used: gyroid (G), gyroid hole (GH), sphere (SP), cylinder (C), slab (S), cylindrical hole (CH), and spherical hole (SH).}
\label{kaiju}
\end{center}
\end{figure}

Nuclei in laboratories are roughly spherical except for some significantly deformed nuclei, which are often encountered in excited states. It is considered, however, that nuclei in matter just below normal nuclear density are not always spherical even in the ground state. Matter in supernova cores and in neutron star crusts are the best candidates where non-excited nuclei that have unusual shapes could be found, surrounded by a gas of dripped neutrons. The idea of non-spherical nuclei in matter was initially introduced when one considered a melting transition of matter with nuclei into uniform nuclear matter as the density increases. It was suggested that before melting, the system turns into a bubble state in which nuclei turn inside out \cite{bbp71, lamb78}. In the subsequent works using liquid drop models (LDMs), it was concluded that nuclei deform from sphere (SP) to cylinder (C), slab (S), cylindrical hole (CH), and spherical hole (SH) before melting into uniform matter \cite{rave83, hashi84, oyak84}. Since such non-spherical nuclei look like spaghetti, lasagna, macaroni, and Swiss cheese, respectively, they are often referred to as nuclear pasta (for reviews, see, e.g., Refs.~\cite{pethi95, nbsnd05, cham08}).  Recently, by analogy with the polymer system, it was suggested that nuclei with a complicated shape called gyroid (G) may appear between the C and S phases, and similarly, the hole structure of gyroid (GH) may appear between the S and CH phases \cite{self09}. The nuclear shapes mentioned here are summarized in Fig.~\ref{kaiju}. 

The possible presence of pasta nuclei in supernova cores and neutron star crusts would play an important role in some astrophysical phenomena. For instance, pulsar glitches, which are thought to be caused by catastrophic unpinning of quantized neutron vortices in the inner crust of a neutron star, could be strongly affected by the configuration of nuclei in an amorphous solid that constitutes the crust \cite{link09}. It remains to be clarified how neutron vortices could pin in such an amorphous solid, particularly in the presence of pasta nuclei.\footnote{Possible relevance of pasta nuclei to the vortex pinning was first proposed in Ref.~\cite{moti97} in terms of pinning-induced nuclear rods. Subsequently, Jones \cite{jones02} concluded from the argument of formation enthalpy that pinning-induced nuclear rods are unlikely to form.} In the context of core collapse supernova explosions and protoneutron star cooling, which are mainly controlled by the neutrino opacity, the possible influence of pasta nuclei on the neutrino opacity needs to be taken seriously. The cross section for the neutrino-nucleus scattering depends on the structure of nuclei \cite{horow04a, horow04b, sonod07}. The neutrino scattering processes are no longer coherent in the directions in which nonspherical nuclei are elongated. It is different from the case of roughly spherical nuclei whose finiteness in any direction leads to constructive interference in the scattering.

Earlier investigations of pasta nuclei basically focus on the zero-temperature phase diagram including the SP, C, S, CH, SH, and uniform phases. The Coulomb and surface energies are calculated in the LDM approach because their delicate competition is responsible for the presence of pasta nuclei at sufficiently small internuclear spacings (e.g., Ref.~\cite{oyak84}). In particular, the pasta phases inevitably appear between the SP and uniform phases at zero temperature in the incompressible limit where the density in the nucleus is fixed at the saturation density. Nevertheless, whether pasta nuclei actually occur in stellar matter remains to be answered. A compressible type of LDMs in which matter in the nucleus is allowed to have various densities helps to answer this question.  Many calculations based on the compressible LDMs were performed by calculating the bulk and surface energies from a single model for the effective nucleon-nucleon interactions (e.g., Ref.~\cite{loren93}) and gave different predictions of the density region of pasta nuclei in neutron star crusts. In fact, such differences are shown to come mainly from the still uncertain density dependence of the symmetry energy \cite{oi07}. In the case of supernova cores where the system is less neutronrich but hotter than matter in neutron star crusts, the density region of pasta nuclei is predicted to be relatively wide at sufficiently low temperatures, while the description of the temperature-induced melting of pasta nuclei is still model dependent.

In calculating the electrostatic energy of matter containing pasta nuclei within the LDMs, one typically assumes a geometrical shape of nuclei and uses the one-dimensional Wigner-Seitz (WS) approximation. For further calculations on the structure of pasta nuclei and a gas of dripped neutrons, one often goes beyond a simple liquid-drop picture and utilizes the Hartree-Fock and Thomas-Fermi methods within the WS approximation (e.g., Refs.~\cite{bv81, ogas82, oyak93}). Beyond the WS approximation, shell effects on bound and unbound nucleons were considered \cite{oyak94, magi02, cham05}; the shell effects on unbound nucleons are often referred to as band effects. On the other hand, multidimensional computations without any assumption of geometrical shape were also performed within the framework of the Thomas-Fermi method \cite{koon85}, the quantum molecular dynamics (QMD) method \cite{maru98, sonod08, gwata09}, and the Hartree-Fock theory \cite{magi02, gogel07, newt09, sebi09}, which reproduce the five shapes listed above. Moreover, some of these multidimensional computations suggest more complex structures. We remark that pairing effects \cite{gogel07}, electron-screening effects \cite{nbiid03, maru05}, and fluctuation-induced displacements of pasta nuclei \cite{gwata00,gwata01,gwata03} have also been taken into account in describing the structure of pasta nuclei.

It is noteworthy that similar phase diagrams are obtained for nanostructures of block copolymers \cite{bafre99}. Block copolymers are made up of two joined chemically distinct polymer blocks and each polymer block consists of a linear series of identical monomers. Since the same polymer blocks assemble spontaneously, phase separation of the domain structure occurs. The domains have a proper size and can not grow larger and larger because the distinct polymer blocks are joined strongly in each block copolymer. A resulting rubberlike entropic restoring force is counterbalanced by the thermodynamic interfacial tension driving macroscopic phase separation. This is a great contrast to the case of pasta nuclei where the Coulomb repulsion is essential to clustering with proper sizes as will be shown later. The shape of polymer domains depends on the fraction of two blocks while the shape of pasta nuclei depends on the volume fraction of the nuclear matter part.

One of the shapes of polymer domains is the gyroid, which is a periodic bicontinuous morphology discovered experimentally for the regions between the C and S counterparts and the S and CH counterparts. Interestingly, in the case of nuclear pasta, some phases of complicated structure are predicted to appear in the corresponding regions by the QMD simulations and Hartree-Fock calculations. Matsuzaki \cite{matsuz06} pointed out the possibility that pasta nuclei may have such morphologies as observed in the polymer systems\footnote{Instead of the gyroid morphology, he studied the double-diamond morphology, which is another periodic bicontinuous structure. Historically, the double-diamond morphology was initially considered to be the most likely periodic bicontinuous structure for the polymer system.} but made estimates that were too rough to address the tiny energy differences between morphologies. According to the precise Coulomb and surface energy evaluations within the LDM, it was found that the G (GH) phase does not appear in the ground state for any density but the energy difference from the most stable phase becomes quite small near the transition point from the C (S) phase to the S (CH) phase \cite{self09}. Furthermore, the volume fraction of nuclei at this point was shown to be $\sim0.35$ (0.65), and this value is very close to the two-block fraction where the corresponding transition occurs in the polymer systems.

The LDM approach to inhomogeneous matter at subnuclear densities was frequently utilized since the earliest studies \cite{lamb78, rave83, hashi84, oyak84}. It has advantages over other approaches not only because calculations of various thermodynamic quantities can be performed more straightforwardly, but also because various corrections can be added to a Weizs{\" a}cker-Bethe mass formula in a systematic manner. One such correction is the curvature correction to the surface energy. The curvature effects, which are implicitly taken into account in the Thomas-Fermi and Hartree-Fock approaches, are often ignored in the LDM approach, but can be taken into account once the curvature coefficient is given. The importance of the curvature energy for nonspherical nuclei was pointed out by Pethick {\it et al.} \cite{pethi83}. They introduced this term so as to explain the difference of the nucleus-bubble transition density between Hartree-Fock and LDM calculations. They noted that, since the sign of the curvature corrections for hole nuclei becomes negative, the curvature corrections act to destabilize the phase with nuclei as compared with that with holes.  In Ref.~\cite{koleh85}, the curvature coefficient was obtained by calculating the curvature thermodynamic potential in the Thomas-Fermi approximation with Skyrme interactions in a manner consistent with calculations of the surface tension.  It was also shown that there are large variations in the curvature energy among the results for different interactions.  Incidentally, the curvature term of the LDM is poorly known even from empirical nuclear masses \cite{seehow75, dm77}.  The curvature corrections for other pasta phases were also investigated by utilizing the Hartree-Fock theory \cite{loren91} and the Thomas-Fermi approximation \cite{douch00}.

In this paper, we build the curvature correction into the energy of pasta nuclei including the gyroid structure as derived within the LDM in Ref.~\cite{self09} and examine the associated change in the transition densities between different shapes. In this paper, we will also address the question of whether or not the gyroid phases are energetically favored by newly allowing for curvature effects. In addition, we will present details of the calculations of the surface and Coulomb terms omitted for want of space in Ref.~\cite{self09}.

It is also interesting to examine the possibility that the transitions of pasta configurations are smooth as proposed by Ravenhall {\it et al.} \cite{rave83}. They regarded sphere, cylinder and slab as three-, two- and one-dimensional spheres, respectively, and extended the surface and Coulomb energy expression derived for these three specific cases to a geometry of general dimensionality $d$. By assuming $d$ to be a continuous variable, they obtained a phase diagram in which the optimal value of $d$ changes continuously with density up to a melting point into uniform matter. In this paper, we also discuss whether the gyroid, the recently proposed nuclear pasta, can be interpreted as a non-integer-dimensional sphere by comparing the Coulomb, the surface, and the curvature energies of the gyroid phase with those of a geometry of general dimensionality $d$.

This paper is organized as follows. In Sec.~\ref{form}, we write down the expressions for the energy of pasta nuclei, which include the curvature correction. Section \ref{result} is devoted to illustration of the phase diagram of nuclear pasta, which allows for the possible gyroid phase, and to energy comparison between the gyroid and the general dimensional spheres. Our conclusions are presented in Sec.~\ref{conclu}.

\section{Energy of matter at subnuclear densities} \label{form}

In this section, we summarize the energy expressions that may be obtained for matter at subnuclear densities within the framework of the LDM, which include curvature corrections. We then derive the size equilibrium condition from the energy minimization by treating the curvature term perturbatively.

\subsection{Liquid drop model}

\begin{figure}
\begin{center}
\includegraphics{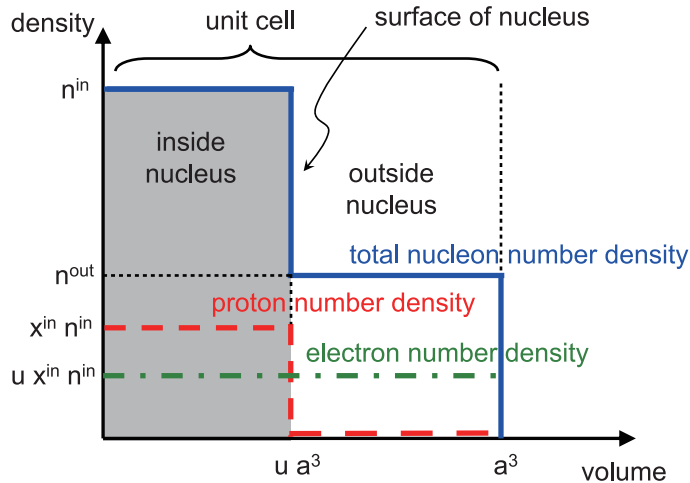}
\caption{(Color online) A schematic density profile in a unit cell as described by the present LDM. The shaded areas in Fig.~\ref{kaiju} correspond to ``inside nucleus'' of this figure.}
\label{cldm}
\end{center}
\end{figure}

In describing the energy of matter at subnuclear densities, we regard nuclei as liquid drops containing neutrons and protons. Here, we deal with the zero-temperature matter, which consists of nuclei with a single shape and size, while nuclei with various shapes and sizes may be mixed at finite temperatures. In our model, nuclei occupy the volume fraction $u$ and have the sharp boundary, uniform number density $n^\mathrm{in}$, and uniform proton fraction $x^\mathrm{in}$. They are embedded in a uniform neutralizing background of electrons of number density $ux^\mathrm{in}n^\mathrm{in}$ and, if any, in a gas of dripped neutrons of uniform number density $n^\mathrm{out}$. For simplicity, we do not take surface diffuseness or neutron skin into account. Since the internuclear Coulomb repulsion gives rise to a spatially periodic structure, we consider a unit cell of volume $a^3$ by setting the shapes of nuclei {\it a priori}. In Fig.~\ref{cldm}, a schematic density profile in a unit cell is shown.

\subsection{Energy expression including curvature term}
We write the energy expression in the form of the semiempirical mass formula which corresponds to the present liquid-drop picture. Including the curvature term and not including rest mass, the total energy of a nucleus with $Z$ protons and $A-Z$ neutrons can be written as
\begin{eqnarray}
E_\mathrm{tot} & = & E_v + E_s + E_\mathrm{curv} + E_\mathrm{sym} + E_\mathrm{ssym} + E_\mathrm{Coul} \nonumber \\
 & = & \left(a_v + a_\mathrm{sym}(1-2x)^2 \right)A 
+ \left(a_s + a_\mathrm{ssym}(1-2x)^2 \right) A^{2/3} 
+ a_\mathrm{curv} A^{1/3} + a_\mathrm{Coul} x^2 A^{5/3},
\label{semiemp}
\end{eqnarray}
where $x=Z/A$. The subscripts $v$, $s$, curv, sym, ssym, and Coul stand for the volume, the surface, the curvature, the symmetry, the surface-symmetry, and the Coulomb terms, respectively. This expression for the curvature term is also seen in Eq.~(1.1) of Ref.~\cite{koleh85}. Correspondingly, we write the total energy of a unit cell $W$ as 
\begin{equation}
W = W_b + W_s + W_{\rm curv} + W_\mathrm{Coul},
\label{totale}
\end{equation}
where $W_b$, $W_s$, $W_\mathrm{curv}$, and $W_\mathrm{Coul}$ are the bulk energy, the surface energy, the curvature energy, and the Coulomb energy, respectively. This expression is the same as Eq.~(3) of Ref.~\cite{self09} except that the curvature energy is introduced. Because of the saturation property of nuclear matter, the mass number $A$ is roughly proportional to the volume of a nucleus. With the volume fraction of a nucleus, $u$, fixed, the volume of a nucleus is proportional to that of the unit cell $a^3$. Therefore, as we will see below, each energy term in Eq.~(\ref{totale}) can be characterized by the $a$ dependence.

Since the bulk energy corresponds to the volume term in Eq.~(\ref{semiemp}), we can write it as
\begin{equation}
W_b = w_b(u, x^\mathrm{in}, n^\mathrm{in}, n^\mathrm{out}) a^3,
\label{bulke}
\end{equation}
with the average bulk energy density $w_b(u, x^\mathrm{in}, n^\mathrm{in}, n^\mathrm{out})$. Note that we encapsulate all the energies that are proportional to the cell volume, such as the electron kinetic energy inside and outside the nucleus, in this term.

The surface energy is proportional to the area of the surface of the nucleus, namely, the interface between a part of nuclear matter and a part of neutron matter. Thus it depends on the shape of the interface, whereas $W_b$ does not. We write the surface energy of a cell as
\begin{equation}
W_s = \sigma(x^\mathrm{in}, n^\mathrm{in}, n^\mathrm{out}) 
      g(u, \mathrm{shape}) a^2,
\label{surfe}
\end{equation}
where $\sigma(x^\mathrm{in}, n^\mathrm{in}, n^\mathrm{out})$ is the surface tension, and the relative surface area $g(u, \mathrm{shape})$ is the area of the surface for $a=1$ with $\mathrm{shape}=\mathrm{SP}$, C, S, CH, SH, G, GH. Generally, it is given by
\begin{equation}
g(u, \mathrm{shape}) = \frac{1}{a^2} \int_{S(u, \mathrm{shape})} \mathrm{d}S,
\label{areaint}
\end{equation}
where the surface integral is taken over the nuclear surface $S(u, \mathrm{shape})$ within a unit cell.

We turn to the curvature energy, which may be expressed as
\begin{equation}
W_{\rm curv} = \omega(x^\mathrm{in}, n^\mathrm{in}, n^\mathrm{out}) 
      h(u, \mathrm{shape}) a,
\label{curve}
\end{equation}
where $\omega(x^\mathrm{in}, n^\mathrm{in}, n^\mathrm{out})$ is the curvature coefficient. $h(u, \mathrm{shape})$ is the integrated mean curvature defined as
\begin{equation}
h(u, \mathrm{shape}) = \frac{1}{a} \int_{S(u, \mathrm{shape})} H(x, y, z)
                       \mathrm{d}S,
\label{curveint}
\end{equation}
where $H(x, y, z)$ is the mean curvature at the point $(x, y, z)$ on $S(u, \mathrm{shape})$. The definition of $H$ is found in textbooks on differential geometry (e.g., Ref.~\cite{text}) [see Eq.~(\ref{donpisha}) for the definition by implicit functions]. Note that $h(u, \mathrm{shape})$ does not depend on $a$ because $H(x, y, z)$ is proportional to $1/a$ and is integrated over the nuclear surface.

The Coulomb energy of a cell is written as
\begin{equation}
W_\mathrm{Coul} = \frac{1}{2} \int_\mathrm{cell} \left[ e \left\{n_p(\bm{r}) - n_e\right\} \phi(\bm{r})\right]\mathrm{d}\bm{r},
\label{coule}
\end{equation}
where $e$ is the elementary charge. $n_p(\bm{r})$ and $n_e$ denote the number densities of protons and electrons, respectively. In our model, their difference is expressed as
\begin{equation}
n_p(\bm{r}) - n_e = \begin{cases}
(1-u) x^\mathrm{in} n^\mathrm{in}, & \text{inside the nucleus,} \\
-u x^\mathrm{in} n^\mathrm{in}, & \text{outside the nucleus.}
\end{cases}
\label{densdiff}
\end{equation}
$\phi(\bm{r})$ is the Coulomb potential determined by the Poisson equation:
\begin{equation}
\nabla^2 \phi(\bm{r}) = - 4 \pi e \left[n_p(\bm{r}) - n_e\right].
\label{pisne}
\end{equation}
We can rewrite Eq.~(\ref{coule}) as
\begin{equation}
W_\mathrm{Coul} = \left(e x^\mathrm{in} n^\mathrm{in} \right)^2 
w_\mathrm{Coul}(u, \mathrm{shape}) a^5,
\label{couler}
\end{equation}
and Eq.\ (\ref{pisne}) as
\begin{equation}
\nabla_s^2 q(\bm{s}) = - 2 \pi p(\bm{s}).
\label{pisner}
\end{equation}
Here, $\bm{s}=\bm{r}/a$ and $\nabla_s$ are the dimensionless vector and its differential operator,
\begin{subequations}
\begin{equation}
w_\mathrm{Coul}(u, \mathrm{shape}) = \int_{\mathrm{cell}^\prime} 
                  p(\bm{s}) q(\bm{s}) \mathrm{d}\bm{s},
\label{scoule}
\end{equation}
\begin{equation}
p(\bm{s}) = \frac{1}{x^\mathrm{in} n^\mathrm{in}} 
             \left[n_p(a\bm{s}) - n_e\right],
\label{scoulp}
\end{equation}
\begin{equation}
q(\bm{s}) = \frac{1}{2e x^\mathrm{in} n^\mathrm{in} a^2} \phi(a\bm{s}),
\label{scoulq}
\end{equation}
\label{scoula}
\end{subequations}
where ``cell$^\prime$'' in the integral denotes the cell normalized by $a^3$. We can see that $W_\mathrm{Coul}(u, \mathrm{shape})$ is proportional to $a^5$, which is consistent with Eq.~(\ref{semiemp}).

\subsection{Energy minimization}

By substituting Eqs.~(\ref{bulke}), (\ref{surfe}), (\ref{curve}), and (\ref{couler}) into Eq.~(\ref{totale}), we rewrite the total energy density as
\begin{eqnarray}
\frac{W}{a^3} & = & w_b(u, x^\mathrm{in}, n^\mathrm{in}, n^\mathrm{out}) 
+ \frac{\sigma(x^\mathrm{in}, n^\mathrm{in}, n^\mathrm{out}) 
   g(u, \mathrm{shape})}{a} 
+ \frac{\omega(x^\mathrm{in}, n^\mathrm{in}, n^\mathrm{out}) 
   h(u, \mathrm{shape})}{a^2} \nonumber \\
 & & + \left(e x^\mathrm{in} n^\mathrm{in} \right)^2 
   w_\mathrm{Coul}(u, \mathrm{shape})a^2.
\label{totaled}
\end{eqnarray}
In the following, we minimize the total energy density. This consists of two steps. The first step is minimization with respect to the size of the unit cell $a$, which leads to
\begin{equation}
\frac{\partial}{\partial a}\left( \frac{W}{a^3} \right) 
= - \frac{\sigma g}{a^2} - \frac{2\omega h}{a^3} 
  + 2\left(e x^\mathrm{in} n^\mathrm{in} \right)^2 w_\mathrm{Coul} a = 0.
\label{minitotaled}
\end{equation}
Since the curvature energy is generally small compared with the surface and Coulomb energies, at first, we solve the equation for $\omega=0$ as 
\begin{equation}
a_0 = \left(\frac{\sigma g}{2(e x^\mathrm{in} n^\mathrm{in})^2 w_\mathrm{Coul}}
      \right)^{1/3}.
\label{a0sol}
\end{equation}
For $\omega h \ll \sigma g a_0$, Eq.~(\ref{minitotaled}) can be solved as
\begin{equation}
a = a_0 \left(1+\frac{2\omega h}{3\sigma g a_0} \right).
\label{asol}
\end{equation}
By eliminating $a$ from Eq.~(\ref{totaled}) and by retaining terms of up to first order in $\omega h / \sigma g a_0$, we obtain
\begin{eqnarray}
\frac{W}{a^3} & = & w_b + 
\frac{3}{\sqrt[3]{4}} \left( e x^\mathrm{in} n^\mathrm{in} \sigma \right)^{2/3}
 g^{2/3} w_\mathrm{Coul}^{1/3} \left( 1 +\frac{2\omega h}{3\sigma g a_0}\right)
  \nonumber \\
 & = & w_b + \frac{3}{\sqrt[3]{4}} \left( e x^\mathrm{in} n^\mathrm{in} \sigma 
\right)^{2/3} g^{2/3} w_\mathrm{Coul}^{1/3} + \sqrt[3]{4} \left[ 
\frac{(e x^\mathrm{in} n^\mathrm{in})^4 \omega^3} {\sigma^2} \right]^{1/3} 
\frac{w_\mathrm{Coul}^{2/3} h}{g^{2/3}}.
\label{totaledagain}
\end{eqnarray}
We remark that we can rewrite Eq.~(\ref{minitotaled}) as $W_s + 2W_{\rm curv} = 2W_\mathrm{Coul}$, an extended version of the well-known condition for size equilibrium, $W_s = 2W_\mathrm{Coul}$, derived for $\omega=0$.

Second, we minimize Eq.~(\ref{totaledagain}) with respect to the shape for given $u$. This is performed by simply comparing the energies for different shapes and by finding the shape that gives the lowest energy. Note that, in Eq.~(\ref{totaledagain}), the shape dependence is entirely encapsulated in the geometrical factors defined as
\begin{subequations}
\begin{equation}
F_0(u, \mathrm{shape}) = g(u, \mathrm{shape})^{2/3} 
                         w_\mathrm{Coul}(u, \mathrm{shape})^{1/3},
\label{fighty0}
\end{equation}
\begin{equation}
F_1(u, \mathrm{shape}) = \frac{w_\mathrm{Coul}(u, \mathrm{shape})^{2/3} 
                         h(u, \mathrm{shape})}{g(u, \mathrm{shape})^{2/3}}.
\label{fighty1}
\end{equation}
\label{fighty}
\end{subequations}
$F_0(u, \mathrm{shape})$ corresponds to the sum of the relative Coulomb and surface energy densities, which was denoted in our previous study \cite{self09} as $F(u, \mathrm{shape})$, whereas $F_1(u, \mathrm{shape})$ corresponds to the relative curvature correction. For $\omega=0$, as already discussed in Ref.~\cite{self09}, the nuclear shape that minimizes the total energy density is determined uniquely for given volume fraction $u$ and is independent of the average bulk energy density $w_b$ and the surface tension $\sigma$. Likewise, the shape dependence of the relative curvature correction is determined uniquely for given $u$. Because of the different shape dependences of $F_0$ and $F_1$, however, the nuclear shape is no longer determined uniquely for given $u$ in the presence of the relative curvature correction.

We remark that we will set specific values for the parameters $n^\mathrm{out}$, $n^\mathrm{in}$, and $x^\mathrm{in}$ by hand, although they can, in principle, be determined by additional energy minimization under boundary conditions associated with electric charge and baryon number. The setting will be made in the spirit of the incompressible LDM.

\section{Equilibrium nuclear shapes} \label{result}

In this section, we seek the equilibrium nuclear shapes by comparing the energy densities of the pasta phases, which include the gyroid phase. In particular, we address the question of whether the effect of the curvature is helpful for the gyroid appearance or not. We also compare the energy of the gyroid phase with that of a generalized dimensional sphere.

\subsection{Candidate shapes}
As usual, we consider sphere, cylinder, slab, cylindrical hole, and spherical hole as candidates for the equilibrium nuclear shape. These shapes are conventional pastas and are regarded as the one-, two-, or three-dimensional spheres that minimize the surface area at constant nuclear volume. In addition, the gyroid phase, the nuclear pasta with a periodic bicontinuous morphology recently proposed by analogy with the nanostructures of block copolymers \cite{self09}, and its hole structure are included in our investigations.
 
For the conventional pasta phases, the relative Coulomb and surface energy densities can be calculated as
\begin{subequations}
\begin{equation}
F_0(u, \mathrm{SP}) = \left( 36 \pi u^2 \right)^{2/9} \left[ \frac{\sqrt[3]{9 \pi}
\left( 2u^{5/3} - 3u^2 + u^{8/3} \right)}{5 \sqrt[3]{2}} + c_\mathrm{bcc}u^2 \right]^{1/3},
\label{spheref}
\end{equation}
\begin{equation}
F_0(u, \mathrm{C}) = \left( 4 \pi u \right)^{1/3} \left[ \frac{u^2}{2} \left( u - 1 - \log u \right) + c_\mathrm{hex}u^2 \right]^{1/3},
\label{clindf}
\end{equation}
\begin{equation}
F_0(u, \mathrm{S}) = \left( \frac{2 \pi}{3} \right)^{1/3} u^{2/3} (1-u)^{2/3}.
\label{sladf}
\end{equation}
\label{pastaf}
\end{subequations}
The numerically determined coefficients, $c_\mathrm{bcc}=6.5620\times10^{-3}$ and $c_\mathrm{hex}=1.2475\times10^{-3}$, are corrections to the WS approximation, and the subscripts bcc and hex represent the body-centered cubic (bcc) and hexagonal (hex) lattices, respectively, which are the most stable alignments for each nuclear shape \cite{oyak84}. For an SHl and a CH, the following relations hold: $F_0(u, \mathrm{CH}) = F_0(1-u, \mathrm{C})$ and $F_0(u, \mathrm{SH}) = F_0(1-u, \mathrm{SP})$ for the reasons as described below. First, the hole morphology with a fraction of $u$ has the same surface area as the normal morphology with a fraction of $1-u$. Thus, the relations such as $g(u, \mathrm{SH})=g(1-u, \mathrm{SP})$ hold. Second, since the charge density (\ref{densdiff}) has a sign opposite to the case of its hole morphology with $u$, the signs of $p(\bm{s})$ defined by Eq.~(\ref{scoulp}) and $q(\bm{s})$ defined by Eq.~(\ref{scoulq}) also become opposite. According to Eq.~(\ref{scoule}), therefore, the relative Coulomb energy satisfies the relation $w_\mathrm{Coul}(1-u, \mathrm{SP})=w_\mathrm{Coul}(u, \mathrm{SH})$.

Similarly, the relative curvature corrections can be obtained as
\begin{subequations}
\begin{equation}
F_1(u, \mathrm{SP}) = \left( \frac{256 \pi^4}{3u} \right)^{1/9} \left[ \frac{\sqrt[3]{9 \pi}
\left( 2u^{5/3} - 3u^2 + u^{8/3} \right)}{5 \sqrt[3]{2}} + c_\mathrm{bcc}u^2 \right]^{2/3},
\label{spheref}
\end{equation}
\begin{equation}
F_1(u, \mathrm{C}) = \left( \frac{\pi^2}{4u} \right)^{1/3} \left[ \frac{u^2}{2} \left( u - 1 - \log u \right) + c_\mathrm{hex}u^2 \right]^{2/3},
\label{clindf}
\end{equation}
\begin{equation}
F_1(u, \mathrm{S}) = 0.
\label{sladf}
\end{equation}
\label{pastaf}
\end{subequations}
For an SH and a CH, $F_1(u, \mathrm{CH}) = -F_1(1-u, \mathrm{C})$ and $F_1(u, \mathrm{SH}) = -F_1(1-u, \mathrm{SP})$ are satisfied for the following reason. While the surface area of the hole morphology with a fraction of $u$ is the same as that of the normal morphology with a fraction of $1-u$, the radii of curvature of the nuclear matter surfaces have signs opposite to each other.  Thus, the relations such as $h(u, \mathrm{SH})=-h(1-u, \mathrm{SP})$ hold. Incidentally, we remark that the volume fraction of the nucleus in a unit cell $u$ has a geometrically allowed range: $u<\sqrt{3} \pi /8$ for an SP with bcc and $u<\pi / 2\sqrt{3}$ for a C with hex.

The gyroid structure is so complicated that we should evaluate $g(u, \mathrm{G})$, $h(u, \mathrm{G})$, and $w_\mathrm{Coul}(u, \mathrm{G})$ numerically. We set the gyroid structure by the following level surfaces:
\begin{equation}
f(x, y, z) = \sin \frac{2 \pi x}{a} \cos \frac{2 \pi y}{a} + \sin \frac{2 \pi y}{a} \cos \frac{2 \pi z}{a} + \sin \frac{2 \pi z}{a} \cos \frac{2 \pi x}{a} = \pm k,
\label{gyroid}
\end{equation}
where $(x, y, z)$ are the spatial coordinates. Here, by taking $a$ as the periodic length, the volume of a unit cube becomes $a^3$. By using this expression, we can assume that the region that satisfies $|f(x, y, z)| > k$ corresponds to the nucleus in Fig.~\ref{cldm}, where $k$ is a positive parameter that specifies the volume fraction $u$ of the nucleus in the unit cube.\footnote{We remark that the surface with $k=0$ is not a minimal surface itself, whose mean curvature vanishes but very close to it. This is because Eq.~(\ref{gyroid}) is just an approximation to the mathematical expression for the gyroid that is originally defined as a family of periodic minimal surfaces. For $u<1$, the equilibrium configurations are most likely characterized by constant but nonzero mean curvature surfaces, which are a more general class of minimal surfaces that are stationary with respect to variations of the surface area for a fixed volume fraction. The mean curvatures of the surfaces (\ref{gyroid}) with corresponding $k$ are not constant; they fluctuate by $\sim$20\% at $u=0.35$. Again, this is caused by the approximation. Nevertheless, we call the surfaces (\ref{gyroid}) ``gyroid'' in this paper.} Note that $k \to 0$ corresponds to $u \to 1$ and that $u$ is a monotonically decreasing function of $k$. One of the notable characters of this structure is that the regions inside and outside the nucleus are bicontinuous. Equation~(\ref{gyroid}) is no longer a good approximation for such small values of $u$ as $u<0.0354$ since the resultant configurations are not bicontinuous but pinched off. Because this poses no problem for the following analysis, we do not consider these configurations. In addition to the gyroid morphology, the hole structure of the gyroid, for which nucleons reside in the region satisfying $|f(x, y, z)| <k$, is taken into account. Again, the configurations are bicontinuous only for $u<0.965$.

To evaluate $g(u, \mathrm{G})$ and $h(u, \mathrm{G})$, we should perform the surface integrals (\ref{areaint}) and (\ref{curveint}) numerically. The mean curvature of a level surface defined by such an implicit function as Eq.~(\ref{gyroid}) is known to be given by
\begin{equation}
H(x, y, z) = \pm \frac{(f_{xx}+f_{yy})f_z^2 + (f_{yy}+f_{zz})f_x^2 + (f_{zz}+f_{xx})f_y^2 -2(f_{xy}f_xf_y + f_{yz}f_yf_z + f_{zx}f_zf_x)}{2(f_x^2+f_y^2+f_z^2)^{3/2}},
\label{donpisha}
\end{equation}
where $f_{xy} = \frac{\partial^2}{\partial x \partial y} f(x, y, z)$ and so on \cite{text}. The sign $\pm$ corresponds to the normal and hole morphologies. Fortunately, in the cases of interest here, we can evaluate the surface integral by converting it to the volume integral via Gauss' divergence theorem. Mathematical details are given in the Appendix. This conversion is useful because the surface integral, in general, is difficult to perform numerically.    

We then evaluate $w_\mathrm{Coul}(u, \mathrm{G})$ by solving the normalized Poisson equation (\ref{pisner}). Here, we utilize the discrete Fourier transform, which is known to be powerful for a periodic cubic box.

\subsection{Curvature effect}
\begin{figure}
\begin{center}
\includegraphics{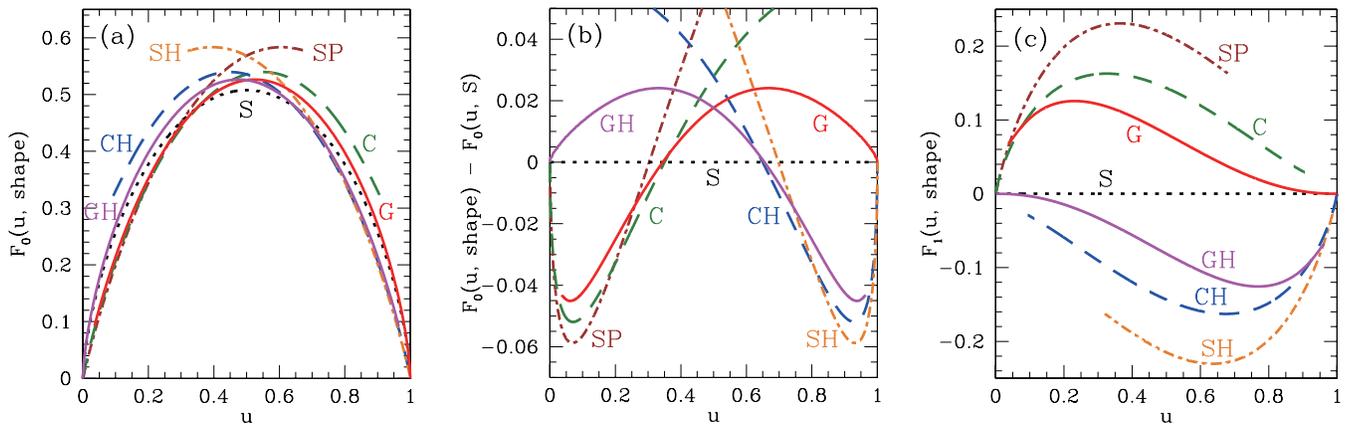}
\caption{(Color online) (a) Relative Coulomb and surface energy densities, (b) their differences from the value of the slab phase, and (c) relative curvature corrections, calculated as functions of $u$ for the seven phases of interest here. The notations are the same as in Fig.~\ref{kaiju}.}
\label{f0f1}
\end{center}
\end{figure}
In Fig.~\ref{f0f1}, we show $F_0(u, \mathrm{shape})$ and $F_1(u, \mathrm{shape})$ as functions of $u$ for all the shapes considered here. The differences from the value of the S phase are also shown for $F_0(u, \mathrm{shape})$. For all the phases except the SP and C phases, $F_0$ and $F_1$ converge to zero at the point $u=1$, which corresponds to uniform matter. Recall that for $\omega=0$, the shape that minimizes $F_0(u, \mathrm{shape})$ is the equilibrium one. Thus, we find that equilibrium nuclei deform from SP to C, S, CH, and SH with increasing $u$. While the G phase does not give the minimum value of $F_0$ for any $u$, the $F_0$ value of the G phase is very close to those of the C and S phases at the transition point from the C phase to the S phase ($u=0.35$). The same is true of the $F_0$ value of the GH phase at the transition point from the S phase to the CH phase ($u=0.65$).

The relative curvature correction is the largest for the SP phase, followed in order by the C, G, S, GH, CH, and SH phases. This hierarchy is due to the following reason. The curvature energy is a correction to the surface energy. Phenomenologically, the surface energy arises because nucleons near the surface have less neighbors to interact with attractively than nucleons near the center. Therefore, the more curved surface inward, the less neighbors exist for nucleons near the surface. This feature indicates that the SP morphology has the largest mean curvature among the phases of interest here for the same volume fraction. The curvature of the S morphology is zero because of its flat surface.  Since the surface of hole nuclei is reentrant, the sign of their curvature corrections becomes negative.

\begin{figure}
\begin{center}
\includegraphics{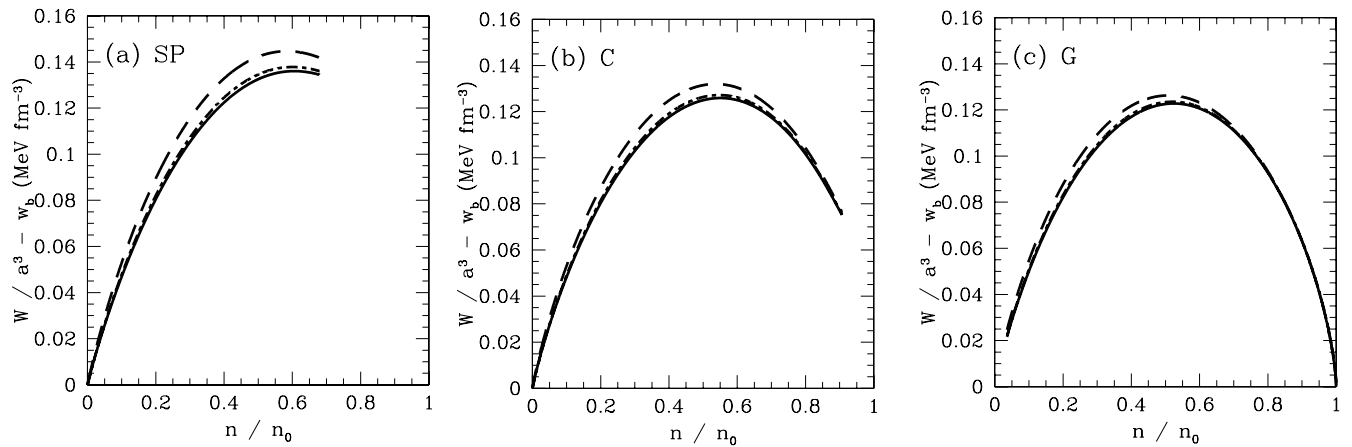}
\caption{Average total energy densities minus the bulk term for the (a) spherical, (b) cylindrical, and (c) gyroid morphologies. The solid, dot-dashed, and dashed lines correspond to the cases for $\omega = 0$~MeV~fm$^{-1}$, $\omega = 0.2$~MeV~fm$^{-1}$, and $\omega = 1$~MeV~fm$^{-1}$, respectively.}
\label{wbara3}
\end{center}
\end{figure}
To determine the nuclear shape that minimizes the total energy density (\ref{totaledagain}) at given $u$, we specify the values of the parameters $n^\mathrm{out}$, $n^\mathrm{in}$, and $x^\mathrm{in}$, which, in turn, are related to the coefficients $w_b$, $\sigma$, and $\omega$. By bearing the application to matter in supernova cores in mind, we follow the setup in the incompressible limit as employed in Ref.~\cite{nbsnd05}, while the formulations and analyses shown above are also applicable to the compressible case. We set $n^\mathrm{out}=0$ and $n^\mathrm{in}=n_0$, where $n_0=0.165$~fm$^{-3}$ is the saturation density. Then, the volume fraction is given by $u=n/n_0$, where $n$ is the average nucleon number density. The proton fraction is set to $x^\mathrm{in}=0.3$. The surface tension is assumed to be $\sigma=0.73$~MeV~fm$^{-2}$, which comes from the liquid-drop parameters $a_s=4\pi R_0^2 \sigma_0$ and $a_{\rm ssym}=-4\pi R_0^2 \sigma_0 C_{\rm sym}$ with $\sigma_0=1$~MeV~fm$^{-2}$, $C_{\rm sym}=1.7$, and $R_0=(3/4\pi n_0)^{1/3}$, which reproduce the properties of isolated finite nuclei in the limit of $u \to 0$. We assume that the curvature coefficient $\omega$ is a constant free parameter because of its uncertainties. In Fig.~\ref{wbara3}, we show the average total energy densities minus the bulk term for several values of $\omega$, which are taken in the range $\omega \le 1$~MeV~fm$^{-1}$ by reference to the calculations from the Skyrme interactions \cite{koleh85,douch00}. We can recognize that the curvature effect on this energy density difference is at most $\sim$10\% and that the assumption that the curvature energy is small is good for $\omega \le 1$~MeV~fm$^{-1}$. Note that we do not have to take a specific value of $w_b$ in the incompressible case as considered here, but $w_b$ is essential for the realistic description of the melting into uniform matter \cite{oi07}.
\begin{figure}
\begin{center}
\includegraphics{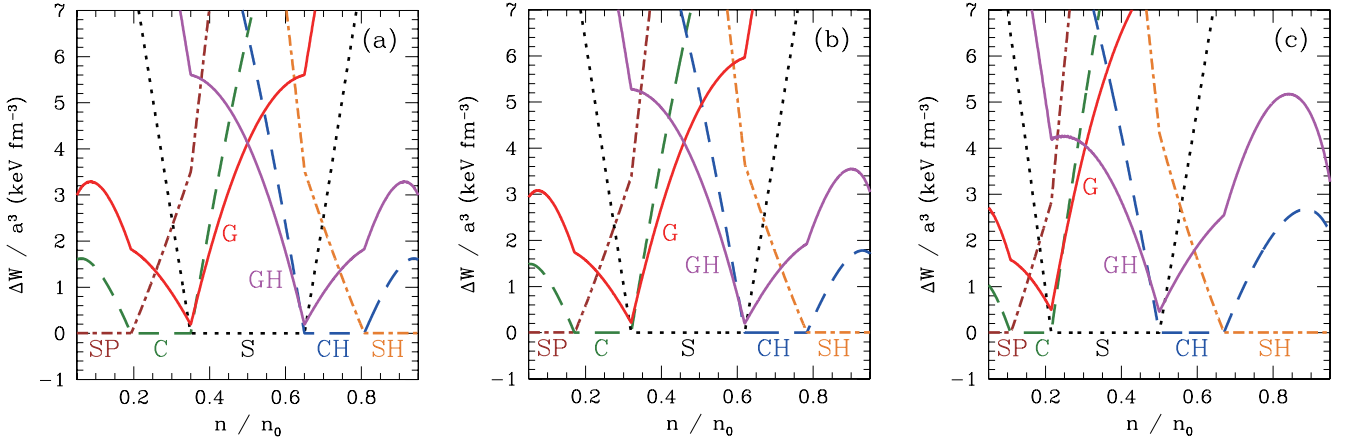}
\caption{(Color online) Energy density differences from the ground state, calculated as functions of $n/n_0$ for (a) $\omega = 0$~MeV~fm$^{-1}$, (b) $\omega = 0.2$~MeV~fm$^{-1}$, and (c) $\omega = 1$~MeV~fm$^{-1}$. The notations are the same as in Fig.~\ref{kaiju}.}
\label{dwbara3}
\end{center}
\end{figure}

\begin{figure}
\begin{center}
\includegraphics{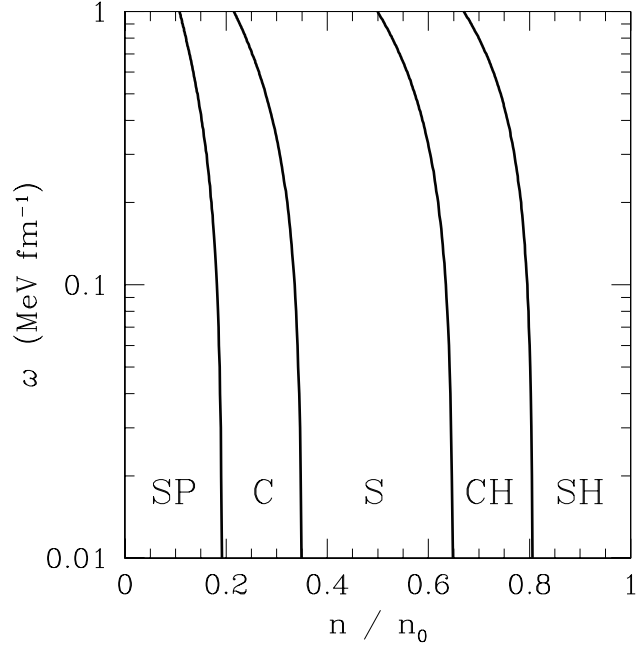}
\caption{Phase diagram on the $n/n_0$ versus the $\omega$ plane, calculated from the LDM with the parameters used in Ref.~\cite{nbsnd05}. The notations are the same as in Fig.~\ref{kaiju}.}
\label{phasedm}
\end{center}
\end{figure}

In Fig.~\ref{dwbara3}, the difference $\Delta W / a^3$ between the average total energy density for each of the shapes considered here and that of the most stable phase at given $n/n_0$ is shown for various values of $\omega$. Whereas we originally expected the G phase to appear as the ground state because of the curvature correction, this is not the case. In the absence of the curvature correction, the total average energy density of the G morphology becomes the same as that of the C morphology at $n/n_0 \sim 0.35$. The density where the C and G morphologies have the same energy is lowered by the curvature correction because the mean curvature of the G morphology is smaller than that of the C morphology. Therefore, the G phase is expected to appear at $n/n_0 \lesssim 0.35$. However, the energy of the S morphology becomes lower than that of the G morphology because the curvature correction acts to raise the energy of the G morphology while keeping that of the S morphology unchanged, which has zero curvature. Thus, the G morphology does not appear for any density. In Fig.~\ref{phasedm}, we show the phase diagram on the $n/n_0$ versus the $\omega$ plane. We can recognize that, for a larger curvature coefficient, the transition densities between different morphologies become lower. Again, the reason is that the mean curvature of the low-density phase is larger than that of the high-density phase.

\subsection{General dimensionality}
In the earliest study on the nuclear pasta \cite{rave83}, the Coulomb plus surface energy of the $d$-dimensional sphere was represented by a single expression, and it was argued that the continuous change of the nuclear shape with density could be characterized by the optimal value of the continuous variable $d$. Here, we investigate whether the gyroid can be regarded as a member of the $d$-dimensional spheres or not. In this subsection, we neglect the corrections to the WS approximation for simplicity. By following a line of argument from Ref.~\cite{rave83}, we write the relative Coulomb and surface energy density $F_0$ and the relative curvature correction $F_1$ for the $d$-dimensional spheres as
\begin{subequations}
\begin{equation}
F_0(u, d) = \left[ \frac{2 \pi d^2 u^3}{d+2} \left( u - \frac{d u^{1-2/d} - 2}{d-2} \right) \right]^{1/3},
\label{f0d}
\end{equation}
\begin{eqnarray}
F_1(u, d) & = & \frac{d-1}{2du} F_0(u, d)^2 \nonumber \\
  & = & \left[ \frac{\pi^2 d (d-1)^3 u^3}{2(d+2)^2} \left( u - \frac{d u^{1-2/d} - 2}{d-2} \right)^2 \right]^{1/3}.
\label{f1d}
\end{eqnarray}
\label{fd}
\end{subequations}
Note that,
\begin{equation}
\lim_{d \to 2} \frac{d u^{1-2/d} - 2}{d-2} = \log u + 1,
\label{dlim2}
\end{equation}
and $d=3$, 2, and 1 correspond to sphere, cylinder, and slab, respectively.

\begin{figure}
\begin{center}
\includegraphics{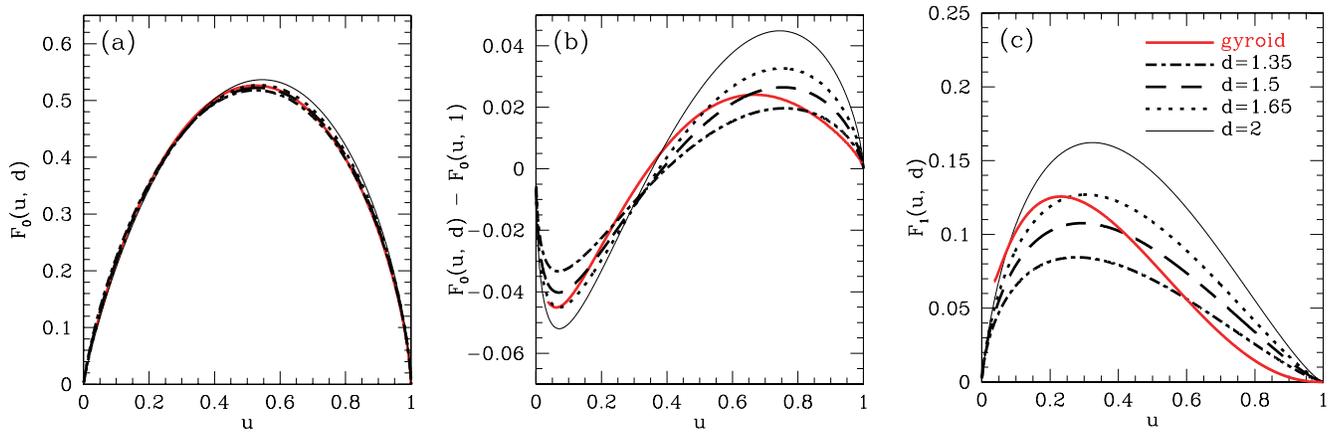}
\caption{(Color online) Same as Fig.~\ref{f0f1} but for the $d$-dimensional spheres and gyroid.}
\label{nonint}
\end{center}
\end{figure}

In Fig.~\ref{nonint}, we compare $F_0$ and $F_1$ of the gyroid with those of the $d$-dimensional spheres with $1<d<2$. Least squares fitting allows us to determine the ``dimension'' of the gyroid as $d=1.478$ for $F_0$ and $d=1.515$ for $F_1$. In this sense, the G phase could be interpreted as the intermediate of the C and S phases. As can be seen in the figure, however, the general dimensionality does not give a good overall fit for $F_0$ and $F_1$ of the gyroid. This fact suggests that the candidates for equilibrium nuclear shape should be examined individually beyond the general dimensionality. 

\section{Conclusions} \label{conclu}
In this paper, we built the curvature correction into the LDM and examined its influence on the equilibrium phase diagram associated with the nuclear pasta including the gyroid structure. We confirmed that the curvature effect pushes the onsets of the shape changes to low densities. We originally expected the gyroid to appear in the phase diagram because of the curvature correction. However, we found that the curvature correction makes the gyroid appearance harder although the effect is not remarkable. We also compared the energy of the gyroid with that of a generalized dimensional sphere. We found that the G phase does not belong to a family of such spheres. This fact implies that the intermediate phases between the conventional pastas should be described beyond the scope of the noninteger dimensionality to know their possible existence in the ground state. 

While we mainly considered supernova matter, the qualitative results for the curvature effect would be unchanged for neutron star matter. In this case, the proton fraction inside a nucleus, $x^\mathrm{in}$, is smaller in the absence of trapped neutrinos, and a gas of dripped neutrons appear in the inner crust of a neutron star. Accordingly, the coefficients $w_b$, $\sigma$, and $\omega$ are greatly modified while $F_0(u, \mathrm{shape})$ and $F_1(u, \mathrm{shape})$, Eq.~(\ref{fighty}), are unchanged. In the present LDM approach, we concluded that the G phase does not appear in the equilibrium phase diagram. However, the energy difference between the G phase and the ground state is so tiny that it would be significant to confirm this conclusion in more sophisticated approaches. 
We itemize various corrections ignored in the present analysis as follows:
(i) Compressibility.  It leads to deviation of the liquid drop density from $n_0$ even in equilibrium, which in turn results in changes in the bulk energy (\ref{bulke}), the surface energy (\ref{surfe}), the curvature energy (\ref{curve}), and the Coulomb energy (\ref{coule}).  To deal with the compressible case, we should assume a specific nucleon-nucleon effective interaction, which gives the density dependence of Eqs.~(\ref{bulke}), (\ref{surfe}), (\ref{curve}), and (\ref{coule}). Also, note that compressibility leads to surface diffuseness and charge screening as categorized below.
(ii) Surface diffuseness.  The nuclear surface is not a sharp boundary because of finite compressibility.  The resultant surface diffuseness corrects the Coulomb energy \cite{bbp71}.
(iii) Neutron skin.  The mean-square radius of the neutron distribution is generally larger than that of the proton distribution, which corrects the bulk energy (\ref{bulke}) \cite{pethi95}.
(iv) Charge screening.  Both protons and electrons redistribute and then contribute to a decrease in the Coulomb energy (\ref{coule}) and an increase in the bulk energy (\ref{bulke}) \cite{bbp71, nbiid03, maru05, myeswi69}.
(v) Coulomb exchange energy.  The proton Fock term corrects the Coulomb energy (8) \cite{bbp71}.
(vi) Thermal fluctuations.  Thermally induced displacements tend to destroy the nuclear pasta structure \cite{gwata00,gwata01,gwata03}.  It is also interesting to ask whether or not the coexistence of the G phase with other phases is possible at finite temperatures.  Note that the G phase tends to appear at finite temperatures for the copolymer systems, because the entropic restoring force, which drives microscopic phase separations, can be dominant over the destructive effect by thermally induced displacements.

In the course of this work, we have devised a systematic method of calculating the Coulomb, surface, and curvature energies of nuclei of various shapes, which include the one that has periodic bicontinuous structure. In light of the fact that pasta nuclei of complicated shapes have recently been taken note of \cite{self09,sonod08,newt09,matsuz06}, we expect that this method could be useful for future theoretical work involved on pasta nuclei.

\begin{acknowledgments}
We are grateful to Shoichi Yamada for fruitful discussions and continuing encouragements. This work was partially supported by the Japan Society for Promotion of Science (JSPS) through Grant No.~21-1189. We acknowledge the hospitality of the Yukawa Institute for Theoretical Physics during the workshop New Frontiers in QCD 2010, where this work was initiated.
\end{acknowledgments}

\appendix

\section{Calculation of the surface integrals} \label{apdx}
\begin{figure}
\begin{center}
\includegraphics[height=50mm]{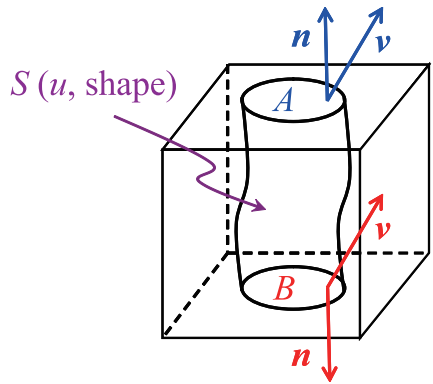}
\caption{(Color online) A schematic of a periodic nuclear surface in a unit cube.}
\label{cancel}
\end{center}
\end{figure}

Equations~(\ref{areaint}) and (\ref{curveint}) include the integrals over the nuclear surface in a unit cell, which are difficult to calculate directly for such complicated surfaces as the gyroid. In this Appendix, we show the numerical method to evaluate these integrals for the gyroid. As mentioned above, a unit cell of the gyroid is a periodic cubic box with volume $a^3$. Thus, we consider a more general surface $S(u, \mathrm{shape})$ and an arbitrary continuously differentiable vector field $\bm{v}(\bm{r})$ that are defined in a unit cube and satisfy the periodic boundary condition. From Gauss' divergence theorem, one can write
\begin{equation}
\oint_S \bm{v}(\bm{r}) \cdot \bm{n}(\bm{r}) \mathrm{d}S = \int_{V(u, \mathrm{shape})} \nabla \cdot \bm{v}(\bm{r}) \mathrm{d}\bm{r},
\label{gauss}
\end{equation}
where $\bm{n}(\bm{r})$ is the outward pointing unit normal vector of the surface $S$ and the volume integral is taken over the region, $V(u, \mathrm{shape})$, surrounded by the surface $S$.  Note that the surface integral in Eq.~(\ref{gauss}) is taken over the cell-nucleus boundaries plus $S(u, \mathrm{shape})$. For simplicity, we consider a case in which the integration region is periodically connected only at the top and bottom sides of the cell as shown in Fig.~\ref{cancel}. Then the integral can be divided as
\begin{equation}
\oint_S \bm{v}(\bm{r}) \cdot \bm{n}(\bm{r}) \mathrm{d}S = \int_{S(u, \mathrm{shape})} \bm{v}(\bm{r}) \cdot \bm{n}(\bm{r}) \mathrm{d}S + \int_A \bm{v}(\bm{r}) \cdot \bm{n}(\bm{r}) \mathrm{d}S + \int_B \bm{v}(\bm{r}) \cdot \bm{n}(\bm{r}) \mathrm{d}S.
\label{surfdiv}
\end{equation}
The integrals over the top side $A$ and bottom side $B$ are canceled out because $\bm{v}(\bm{r})$'s on $A$ and $B$ are parallel because of the periodic condition but $\bm{n}(\bm{r})$'s on $A$ and $B$ are obviously antiparallel. Consequently, only the integral over the nuclear surface $S(u, \mathrm{shape})$ remains. Thus, we get
\begin{equation}
\int_{S(u, \mathrm{shape})} \bm{v}(\bm{r}) \cdot \bm{n}(\bm{r}) \mathrm{d}S = \int_{V(u, \mathrm{shape})} \nabla \cdot \bm{v}(\bm{r}) \mathrm{d}\bm{r}.
\label{s2v}
\end{equation}
Substitution of $\bm{v}(\bm{r})=\bm{n}(\bm{r})$ and $H(\bm{r})\bm{n}(\bm{r})$ into Eq.~(\ref{s2v}) enables us to calculate the integrals in Eqs.~(\ref{areaint}) and (\ref{curveint}), respectively. 

The unit normal vector of a level surface defined by such an implicit function as Eq.~(\ref{gyroid}) is $\pm \nabla f(\bm{r}) / |\nabla f(\bm{r})|$, where the sign $\pm$ corresponds to the inward and outward pointing vectors. Note that there are a finite number of singular points where $\nabla f(\bm{r})=0$. Fortunately, in the present case, these singular points are removable because the singular points are isolated ones rather than lines and planes. First, let us remove small spheres which enclose the singular points from the integration region of the right-hand side of Eq.~(\ref{s2v}). Correspondingly, the integrals over the surfaces of the small voids are added to the left-hand side of Eq.~(\ref{s2v}). These surface integrals vanish in the limit $\varepsilon \to 0$, where $\varepsilon$ is the radius of the removed spheres. This is obvious in the case of $\bm{v}(\bm{r})=\bm{n}(\bm{r})$ because the absolute values of the surface integrals are, at most, $4 \pi \varepsilon^2$. In the case of $\bm{v}(\bm{r})=H(\bm{r}) \bm{n}(\bm{r})$, the surface integrals are $O(\varepsilon)$. Thus, we can evaluate the surface integrals in Eqs.~(\ref{areaint}) and (\ref{curveint}) by numerically computing the volume integrals that have the singular points removed.

\bibliographystyle{apsrev}
\bibliography{apssamp3}

\end{document}